\documentclass[preprint,12pt]{elsarticle}

\usepackage{graphicx}
\usepackage{amssymb}

\usepackage{lineno}

\usepackage{lineno,hyperref}
\usepackage{subfig} 
\usepackage{siunitx} 
\usepackage{booktabs} 
\usepackage{amsmath} 
\usepackage{graphicx, amssymb, wasysym,epsf, graphicx, outlines, enumitem, tikz}
\usepackage{chemformula}
\usepackage{textcomp} 
\graphicspath{{Figures/}}

\journal{Journal of Colloid and Interface Science}

\begin{document}

\begin{frontmatter}


\title{Rheology of protein-stabilised emulsion gels envisioned as composite networks.\\ 2 - Framework for the study of emulsion gels }



\author[Ad:Unilever,Ad:UoE]{Marion Roullet\corref{cor1}}
\ead{marion.roullet@espci.org}
\author[Ad:UoE]{Paul S. Clegg}
\ead{paul.clegg@ed.ac.uk}
\author[Ad:Unilever]{William J. Frith}
\ead{bill.frith@unilever.com}

\address[Ad:Unilever]{Unilever R\&D Colworth, Sharnbrook, Bedford, MK44 1LQ, UK}
\address[Ad:UoE]{School of Physics and Astronomy, University of Edinburgh, Peter Guthrie Tait Road, Edinburgh, EH9 3FD, UK}
\cortext[cor1]{Current address: BioTeam/ECPM-ICPEES, UMR CNRS 7515, Universit\'{e} de Strasbourg, 
	25 rue Becquerel, 67087 Strasbourg Cedex 2, France
}

\begin{abstract}
	\subsection*{Hypothesis}
	The aggregation of protein-stabilised emulsions leads to the formation of emulsion gels. These soft solids may be envisioned as droplet-filled matrices. Here however, it is assumed that protein-coated sub-micron droplets contribute to the network formation in a similar way to proteins. Emulsion gels are thus envisioned as composite networks made of proteins and droplets.
	\subsection*{Experiments}
	Emulsion gels with a wide range of composition are prepared and their viscoelasticity and frequency dependence are measured. Their rheological behaviours are then analysed and compared with the properties of pure gels presented in the first part of this study.
	\subsection*{Findings}
When the concentrations of droplets and protein are expressed as an effective volume fraction, the rheological behaviour of emulsion gels is shown to depend mostly on the total volume fraction, while the composition of the gel indicates its level of similarity with either pure droplet gels or pure protein gels. These results help to form an emerging picture of protein-stabilised emulsion gel as intermediate between droplet and protein gels. This justifies \textit{a posteriori} the hypothesis of composite networks, and opens the road for the formulation of emulsion gels with fine-tuned rheology.
 
\end{abstract}

\begin{keyword}
Colloidal gel \sep Rheology \sep Emulsion \sep Sodium caseinate \sep Viscoelasticity \sep Protein-stabilised droplet \sep Formulation \sep Mixture


\end{keyword}

\end{frontmatter}

\section*{Graphical abstract}
\begin{figure}[htbp]
	\begin{center}
		\includegraphics[width=13cm]{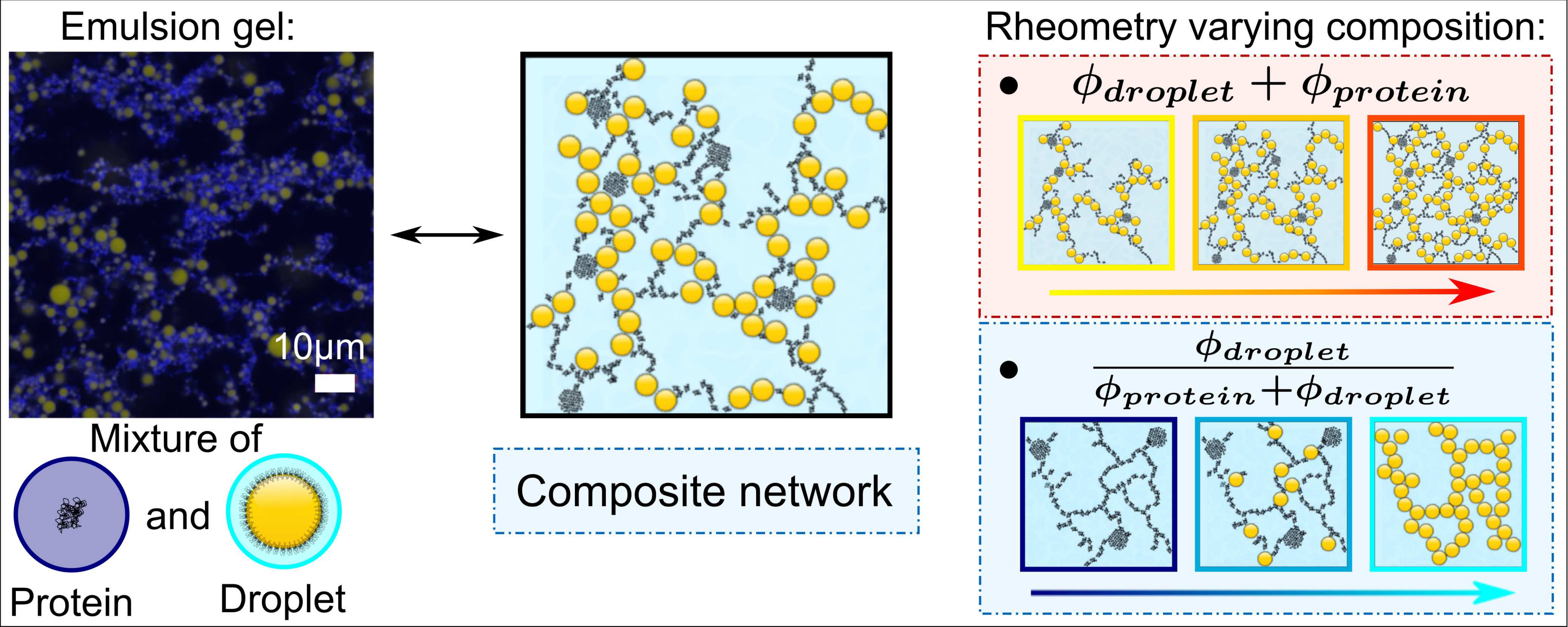}
	\end{center}
	\label{Fig:GraphAbstract}
\end{figure}


\section{Introduction}

Emulsion gels are materials of great interest, because of their many applications in foods, drug-release pharmaceutical products, and novel personal care products  \cite{geremias-andrade:2016,thakur:2012,lupi:2014}. Emulsion gels are soft solids that contain a liquid phase, usually water, trapped within the pores of a network comprised of emulsion droplets \cite{dickinson:2012}. However, this general description conceals the very different structures that emulsion gels can have, depending on their composition \cite{farjami:2019}. Despite the increased efforts in relating the mechanical properties of emulsion gels to their composition, the full understanding of these links is still lacking.

Traditionally, for emulsion gels, the distinction is made between emulsion-filled gels - in which the droplets act as fillers in a viscoelastic gel matrix - and emulsion particulate gels - in which aggregated droplets form a gel network of their own \cite{dickinson:2012,farjami:2019}. Emulsion-filled gels have been studied widely, and a mean field theoretical approach has been used to model the gel matrix, that is often a protein gel, as a continuous medium \cite{vanderpoel:1958,chen:1999a}. In that framework, the emulsion droplets are elastic inclusions that can be deformed upon stressing the emulsion gel \cite{palierne:1990}, and that present interactions with the matrix that are either attractive (active fillers) or repulsive (passive fillers) \cite{vanvliet:1988,chen:1999b,dickinson:1999,koenig:2002,sala:2007,gravelle:2015}. Emulsion particulate gels have attracted less attention, and they were considered to be similar to other colloidal gels \cite{helgeson:2014,krall:1998,zaccarelli:2007}. An exception is the first part of this series, in which pure gels made of protein-stabilised emulsion droplets have been studied and their rheological properties characterised \cite{roullet:2020a}.

At this point, it is useful to note the difference between emulsion gels and concentrated emulsions like mayonnaise, which also display a solid-like behaviour. Emulsion gels present a solid gel network, that can be relatively dilute, and that traps a significant amount of solvent within its pores. By contrast, concentrated emulsions are made of jammed repulsive droplets, that are limited in their mobility by the presence of the other droplets \cite{hemar:2000}. Such jammed systems are often refered to as colloidal glasses \cite{graves:2005}, and are comparable to other glasses made of soft particles, such as star polymers and microgels \cite{mattsson:2009,pellet:2016,conley:2017,gury:2019}. The present study will focus on emulsion gels, and the emulsions used to prepare the gels will thus be kept at concentrations for which a low-shear viscosity can be defined.

The binary distinction between emulsion-filled gels and emulsion particulate gels is however limited by the strong assumptions that are made when defining these two situations. First, the approximation of a continuous matrix, in which the droplets are embedded, does not always apply. Indeed, this matrix is often a protein gel, which is a ramified network with a mesh size of a few microns \cite{mellema:2000,pugnaloni:2005}, so it is assumed that the droplets are much larger. Yet, sub-micron droplets are widely used in commercial products, as their production is made easier by the advances in emulsification techniques, and notably the use of microfluidizers \cite{jafari:2007,schroen:2016}. There can thus be emulsion gels in which the droplets are smaller than the pores of the matrix, which cannot then be approximated by a continuous medium.

Furthermore, it is worth noting that the formation of large aggregates and networks of attractive droplets would make a significant contribution to the overall viscoelasticity of the emulsion gel, but this is generally not considered in the emulsion-filled gel model, while it is central to the existence of emulsion particulate gels. Previous efforts to account for droplet aggregation, and its contribution to the viscoelasticity, in the emulsion-filled gel model, have not yet lead to an accurate estimation of the changes in viscoelasticity induced by droplet aggregation \cite{oliver:2015}. It thus appears necessary to fill the gap between these two models of emulsion gels, to define a more accurate framework for the study of these materials.

This study focuses on the gels produced by destabilising protein-stabilised emulsions, in which proteins - more specifically sodium caseinate - act both as surface-active emulsifier, to form sub-micron oil droplets stabilised by steric and electrostatic repulsion at neutral pH, and as gelling agent. When the emulsion is acidified, the electrostatic repulsion is decreased, and at the protein isoelectric point, attractive van der Waals interactions lead to the gelation of the proteins and of the protein-coated droplets \cite{chen:1999a}. In order to ensure a sufficient surface coverage of the droplets in real systems, and thus a good stability of protein-stabilised emulsions, it is common to work with a protein excess, so a mixture of protein-coated droplets and of unadsorbed proteins is obtained after emulsification \cite{srinivasan:1999}. In summary, the gels studied here are made of sub-micron droplets covered with proteins, and of sodium caseinate, structured into self-assembling aggregates of around $\SI{20}{\nano\meter}$. The oil droplets are part of the network, as they exhibit an attractive interaction between them mediated by the adsorbed proteins at their interface. The protein-stabilised droplets and caseinate assemblies presented here have been thoroughly characterised in a previous study \cite{roullet:2019a}.

In the first paper of this pair, pure gels of caseinate assemblies and pure gels of caseinate-stabilised sub-micron droplets were prepared and characterised \cite{roullet:2020a}. It was shown that the gelation of protein suspensions and of purified suspensions of droplets led to gel networks with a characteristic length-scale of the order of a few microns. The emulsions studied here are thus characterised by droplets that are smaller than the length-scale of the networks, and these droplets aggregate extensively to form a space-spanning fractal network, even at low concentration.

In addition, it was shown that the concentrations of the suspensions of proteins, and of protein-stabilised droplets, could be scaled by the effective volume fraction $\phi_{eff}$, and their viscosity could be analysed in the framework developed for soft colloids \cite{roullet:2019a}. This parameter $\phi_{eff}$ represents the volume occupied by the particles in the sample divided by the total volume. It is calculated by multiplying the concentration by a parameter $k_0$, derived by approximating protein aggregates and protein-stabilised droplets to model hard spheres when in semi-dilute suspensions. This same framework was used to study the gels formed by the two types of suspensions in the first part of this series, and the scaling by the effective volume fraction $\phi_{eff}$ made it possible to emphasise the similarities between the two types of gels at fixed $\phi_{eff}$, both in terms of microstructural features and of rheological properties \cite{roullet:2020a}.

The present work envisions protein-stabilised emulsion gels as composite networks made of un-adsorbed protein assemblies and protein-coated droplets. This approach relies on the hypothesis that there is little distinction between droplets and un-adsorbed proteins in the way each contributes to the properties of the gel of mixture. This is because the most relevant length-scale to study the rheological and microstructural features of colloidal gels appears to be the length-scale of strands of particles \cite{colombo:2013,colombo:2014b,bouzid:2018,delgado:2016}. Consequently, the systems are examined at a much larger scale than of the single particles, and the discrepancy in size and structure of the protein aggregates and protein-stabilised droplets is thus assumed not to be critical. 

Here, protein-stabilised emulsion gels with a wide range of protein and droplet content are prepared, and their rheological properties are characterised and analysed as a function of the composition of the sample. The emulsion gels are then compared to the pure gels of proteins and droplets, to identify the contribution of each of the components to the rheological properties of the composite networks. The emulsion gels are shown to display an intermediate behaviour between those of pure gels of proteins and pure gels of droplets, thus confirming that the framework of composite networks is more relevant for these systems than the two models identified in the existing literature.

\section{Materials \& Methods}

	\subsection{Preparation of protein and droplet suspensions}
Suspensions of pure sodium caseinate and of pure sodium caseinate-stabilised droplets were prepared as described in a previous study \cite{roullet:2019a}, at a range of concentrations.
They were then used as sols for the preparation of acid-induced gels.

\subsection{Preparation of the mixtures of proteins and droplets}

Sodium-caseinate emulsions of well-characterised compositions were prepared by mixing precise amounts of the protein suspension and of the paste of purified droplets. A wide range of compositions of mixtures was explored. In the following, the terms mixture and emulsion will be used without distinction to indicate the samples prepared in this section (as opposed to a standard emulsion where the amount of un-adsorbed protein is uncontrolled).

\subsubsection{Preparation protocol}

To prepare emulsions with a controlled amount of proteins in suspension, the paste of purified droplets at  $\SI[separate-uncertainty=true]{0.519(0008)}{\gram\per\milli\liter}$ was re-suspended in a protein suspension. The protein suspension was prepared as described previously at $\SI{45}{\milli\gram\per\milli\liter}$ and diluted to the desired concentration. As for the suspensions of pure droplets, the paste was first  roughly homogenised with a spatula in the vial, and then gently mixed using a magnetic stirrer. The mixing time required to obtain a visibly homogeneous sample ranged from $\SI{5}{\minute}$ to $\SI{2}{\hour}$. The re-dispersion required longer stirring times at high concentration of droplets and at high concentration of proteins. At a given droplet concentration, significantly more stirring was required to disperse the droplets in a protein suspension than in water.

\subsubsection{Composition of the mixtures}

It is useful to think about protein-stabilised emulsions as ternary mixtures, made of water and of two sorts of colloidal particles: droplets and protein aggregates. Table S1, in the supplementary material, gives the composition of all the mixture samples. The concentrations were calculated from the dilution of the stocks of pure proteins and pure droplets, while volume fractions were calculated from the concentrations as detailed in a previous study \cite{roullet:2019a}.

	\subsection{Preparation of emulsion gels}
	The gels were prepared as described in the first part of this study \cite{roullet:2020a}. 
	The decrease in pH required for the gelation of the sols to occur was induced by the slow hydrolysis of glucono $\delta$-lactone (Roquette).
	The amount of glucono $\delta$-lactone was calculated appropriately for the protein and droplet contents of each sample, using the following weight ratios: for protein $\frac{glucono~\delta-lactone}{protein}=\num{0.185}$, and for caseinate-stabilised droplets, $\frac{glucono~\delta-lactone}{droplet}=\num{0.075}$. The final pH was kept between $\num{4.5}$ and $\num{5.0}$ as in this range of pH, caseinate is at its isoelectric point and thus forms strong gels \cite{chen:1999a}.
	
	The gelation was performed at $\SI{35}{\celsius}$, in order to accelerate the phase transition. Indeed, over long time scales, adverse phenomena such as creaming or bacterial growth may occur in the samples. Following this protocol, the gelation of the suspensions takes between $\SI{30}{\minute}$ and $\SI{2}{\hour}$ , depending on the type of sample and concentration.
	
	The sols containing glucono $\delta$-lactone were placed in the rheometer cup just after preparation, and the measurements were started immediately.

	\subsection{Rheological measurements}
	
Oscillatory rheology measurements were performed using a stress-controlled MCR 502 rheometer (Anton Paar) and a Couette geometry ($\SI{17}{\milli\metre}$ diameter profiled bob and cup CC17-P6, inner radius $\SI{16.66}{\milli\meter}$, outer diameter $\SI{18.08}{\milli\meter}$ yielding a $\SI{0.71}{\milli\meter}$ tool gap, gap length $\SI{25}{\milli\meter}$). To avoid slip at the wall during shearing, profiled bob and cup (serration width $\SI{1.5}{\milli\meter}$, serration depth $\SI{0.5}{\milli\meter}$) were selected as measurement tools.
The temperature was set by a Peltier cell at $\SI{35}{\celsius}$ during the entire measurement sequence. To prevent evaporation, a thin layer of silicon oil of low viscosity ($10$ cSt) was deposited on the surface of the sample.
	
	The measurements were started immediately after mixing of the sample with glucono $\delta$- lactone and subsequent loading in the instrument. First, small-amplitude oscillations (superposition of sinusoids of amplitude $\gamma_0=\SI{0.5}{\percent}$ at the frequencies $\SI{0.2}{\hertz}$, $\SI{0.6}{\hertz}$, $\SI{1}{\hertz}$, $\SI{2}{\hertz}$, $\SI{5}{\hertz}$, $\SI{10}{\hertz}$, with a maximum amplitude of $\gamma_0=\SI{4.0}{\percent}$) were applied during $\SI{9000}{\sec}$ to follow the development of viscoelasticity with time during gelation. Then a frequency sweep was applied to measure the frequency dependence of the moduli for the newly formed gel: with the multiwave mode still activated, the frequency was logarithmically increased ($f= \num{0.005} \ldots \SI{50}{\hertz}$ for the base frequency) at fixed amplitude ($\gamma_0=\SI{4}{\percent}$).
For each sample, $3$ measurements were performed and the values were averaged.

\section{Results \& Discussion}


\subsection{Description of gels of mixtures: choice of the composition parameters}

In order to achieve a thorough study of emulsion gels, it is important to study the full range of what is described as an emulsion, and thus to vary the contents of droplets and un-adsorbed proteins, both in terms of the total concentration and also the ratio of the two components. The choice of parameters for the composition of the gels is a core part of the framework applied to the problem of the study of mixtures.

Previous studies of emulsion gels have focused on the contribution of the droplets \cite{rosa:2006,koenig:2002}, or of the matrix \cite{lupi:2014} to the properties of the gels. Because the role of these two components were considered distinct and studied separately, the individual concentrations were used in the literature to describe the composition of the gels.

\begin{figure}[htbp]
	\begin{center}
		\includegraphics[width=0.9\textwidth]{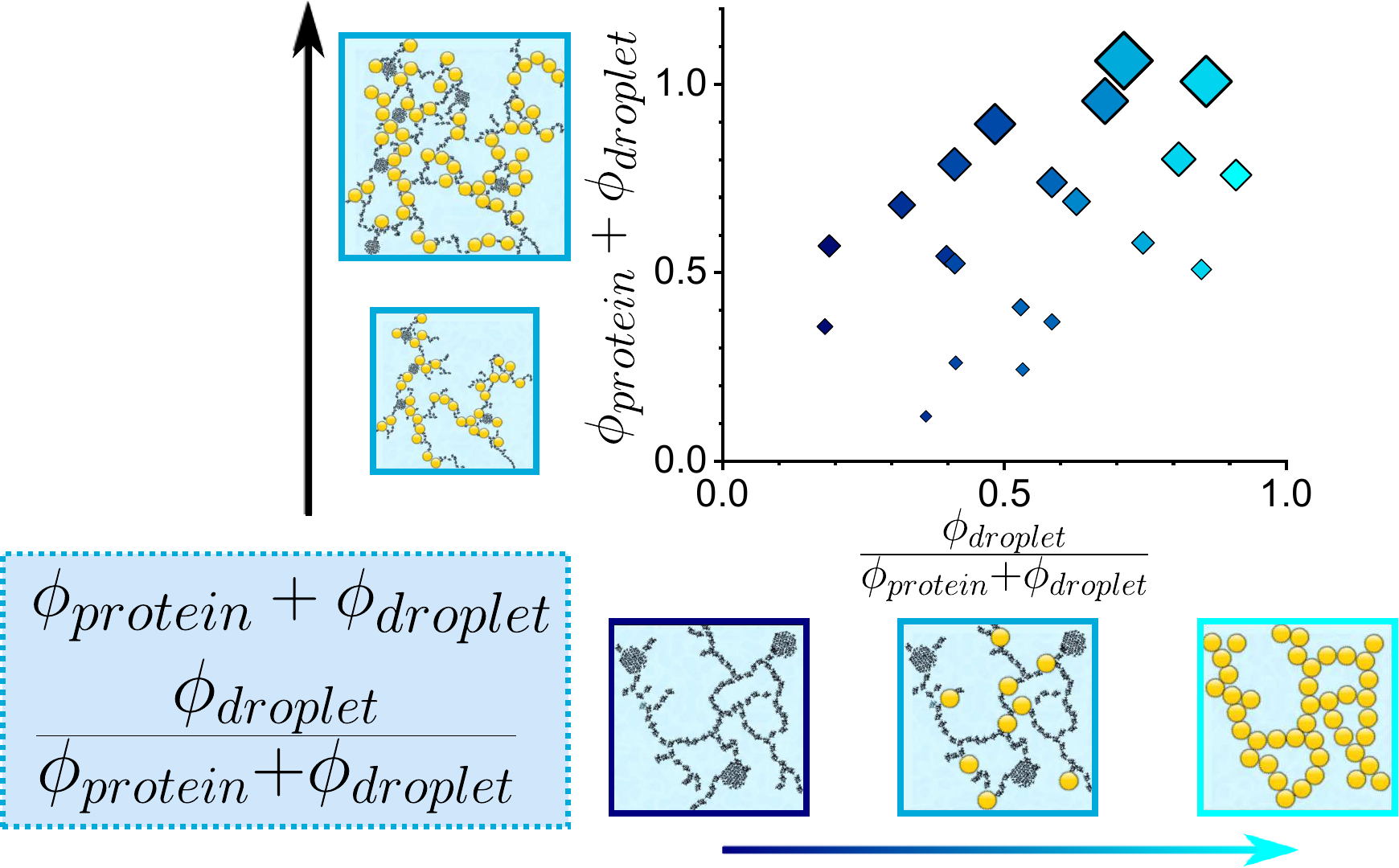}
	\end{center}
	\caption{Composition of the emulsion gel samples prepared in this study. The gels are envisioned as composite networks, and thus described by their total volume fraction $\phi_{eff, prot}+\phi_{eff, drop}$ (coded by the size of the symbols) as a function of the ratio of droplets over the total volume fraction  $\phi_{eff, drop}/(\phi_{eff, prot}+\phi_{eff, drop})$ (colour coded).
	}
	\label{Fig:ParamMixt}
\end{figure}

However, in the present study, the emulsion gels are envisioned as composite networks, similar to the gels formed by the pure gels made of proteins or of droplets. Thus, in order to make possible the comparison of emulsion gels with pure gels,  the focus of the new framework has to be changed, from the individual content of each component in the mixture $\phi_{eff, prot}$ and $\phi_{eff, drop}$, to the total content $\phi_{eff, tot}=\phi_{eff, prot}+\phi_{eff, drop}$ and their relative amounts, described here as $\phi_{eff, drop}/(\phi_{eff, prot}+\phi_{eff, drop})$. This choice of parameters is presented in Figure~\ref{Fig:ParamMixt}. 


As can be seen, these two parameters make the distinction between gels that are similar to protein gels ($\phi_{eff, drop}/(\phi_{eff, prot}+\phi_{eff, drop})\simeq 0$) and gels that are closer to droplet gels ($\phi_{eff, drop}/(\phi_{eff, prot}+\phi_{eff, drop})\simeq 1$), as well as between sparse gels and very dense gels. Their choice thus makes it possible to compare the gels containing a mixture of proteins and droplets to the pure gels of each component. Such a comparison relies heavily on the previous part of this study, in which the rheological properties of pure droplet gels and of protein gels have been characterised over a range of volume fraction.

A general understanding of the properties of emulsion gels can only be built on the exploration of their bi-dimensional composition range. Indeed, such a study allows discrimination of the rheological properties arising from the particle content of the gel, from those related to the composition of the mixture.

It has to be noted that the choice of variables presented in Figure~\ref{Fig:ParamMixt} is not a mere representation tool, but an essential ingredient for the analysis of the rheology of emulsion gels that embeds the vision of these systems as composite networks. The relevance of this choice will be discussed later in light of another framework commonly used for emulsion gels.

\subsection{Rheology of gels of mixtures}


\subsubsection{Viscoelastic properties: decoupling of total volume fraction and composition}

\begin{figure}[htbp]
	\begin{center}
		\includegraphics[width=0.75\textwidth]{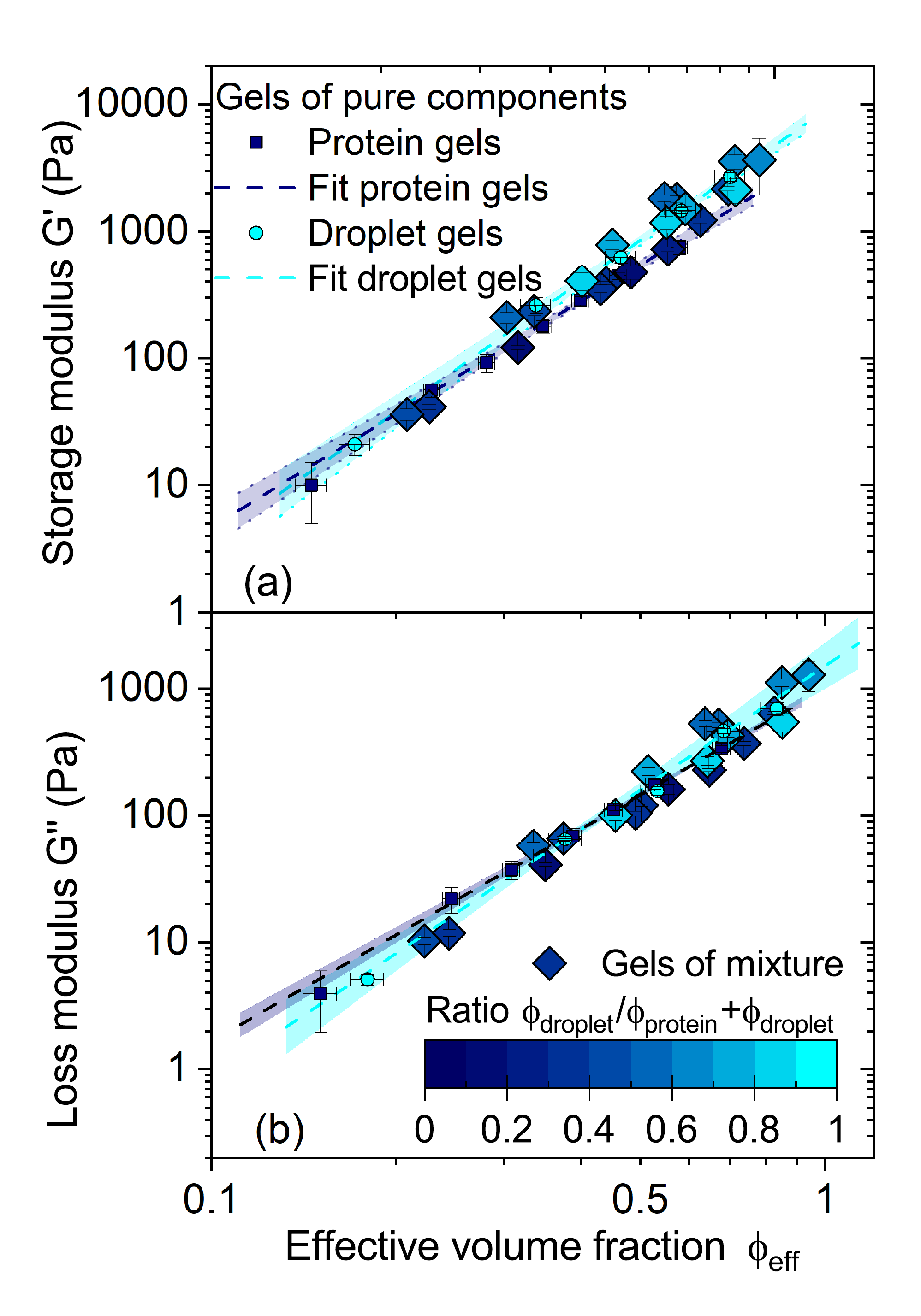}
	\end{center}
	\caption{Storage ($G'$, (a)) and loss ($G''$, (b)) moduli at $\SI{1}{\hertz}$  of protein-stabilised droplet gels (circles, cyan), of protein gels (squares, navy blue), and of gels of mixtures (diamond, colour-coded by the value of $\phi_{eff, drop}/(\phi_{eff, prot}+\phi_{eff, drop})$) as functions of the effective volume fraction of the gel (respectively $\phi_{eff,drop}$, $\phi_{eff,prot}$ and $\phi_{eff,prot}+\phi_{eff,drop}$, scaling derived in Ref.\cite{roullet:2019a}). A power-law fit was performed for each system in the first part of this study, and the model as well as the $\SI{95}{\percent}$ confidence band are displayed on each graph. The horizontal and vertical error bars are calculated using the error propagation theory.
	}
	\label{Fig:GelMixtModuli}
\end{figure}

The rheological properties of the emulsion gel samples, whose compositions are presented in Figure~\ref{Fig:ParamMixt}, were measured during and after gelation. To compare the viscoelasticity of the gels at similar ageing state, the differences in gelation kinetics between samples were taken into account following the same protocol as in the first part of this series \cite{roullet:2020a}. In short, the gelation curves were first shifted horizontally and vertically in logarithmic scale to achieve collapse onto a master curve \cite{ruis:2007,meunier:1999,calvet:2004}. The storage and loss moduli were then measured at a given scaled time on the master curve, as presented In Figure S1 of the supplementary material. The storage and loss moduli of emulsion gels can be compared with the moduli of the gels of pure components presented in the first part of this study \cite{roullet:2020a}. These results are displayed in Figure~\ref{Fig:GelMixtModuli}.

As can be seen, the moduli of the emulsion gels are of the same order of magnitude as for the pure gels and they follow the same trend with the volume fraction. The elastic and viscous aspects of the network are thus mainly determined by the total effective volume fraction $\phi_{eff,droplet}+\phi_{eff,protein}$, and only moderately by the composition. As can be seen in Figure S2 in the supplementary material, this finding is not visible when the weight concentration is used. This result demonstrates that the use of the effective volume fraction developed for suspensions of pure components in Ref.\cite{roullet:2019a} is also relevant for the description of emulsion gels, despite the approximations used.

In addition, small variations in the viscoelastic properties of emulsion gel samples with similar volume fractions but different compositions seem to imply that the nature of the elementary particles forming the network must be taken into account for a more detailed description. Two different approaches for the analysis of the storage moduli, shown in Figure~\ref{Fig:GelMixtModuli} (a), are therefore suggested here to emphasise the influence of the composition and the reinforcement of the gels.

\subsubsection{Reinforcement of gels by fillers: symmetry of components}

First, the classical droplet-filled gel approach can be used for the analysis of the influence of the composition of emulsion gels on their viscoelasticity. In this way, emulsion gels can be considered either as protein gel matrices filled with droplets, or as droplet gel matrices filled with proteins. In this framework, it is interesting to look at the change in rheological properties of the matrix gel upon addition of fillers. The presence of attractive van der Waals interactions between protein-stabilised droplets and proteins when gelation occurs indicates that the addition of filler probably has a reinforcing effect \cite{vanvliet:1988}. 

This reinforcing effect of the component chosen as filler, droplets for example, on the strength of the matrix of the other component, here the protein gel, can be expressed by the ratio of storage moduli between mixture and matrix:
\begin{equation}
\frac{G'^{~exp}_{mixture}}{G'^{~model}_{protein}(\phi_{protein})}
\label{Eq:ReinforcMixt}
\end{equation}
Where $G'^{~exp}_{mixture}$ is the experimental storage modulus of the mixture, as presented in Figure~\ref{Fig:GelMixtModuli}. $G'^{~model}_{protein}$ is the modulus of a hypothetical protein gel, containing the same volume fraction of protein $\phi_{protein}$ as the mixture, and calculated using the model developed in the first part of this series:
\begin{equation}
G'(\phi_{eff}) = G'_{0,\phi} \times \phi_{eff}^{\alpha}
\label{Eq:PureGels}
\end{equation}
The values of the parameters $G'_{0,\phi}$ and $\alpha$ found in the first part of the series are summarised in Table~\ref{Tab:PowerLaw} \cite{roullet:2020a}.

\begin{table}[hbt]
	\centering
	\caption{Parameters to calculate $G'^{~model}_{protein}$ and $G'^{~model}_{droplet}$ using Equation~\ref{Eq:PureGels}.}
	\begin{tabular}{|l|cc|}
		\hline
		Gel type	&  $G'_{0,\phi}$  & $\alpha$ \\
		\hline
		Droplet gels & \SI[separate-uncertainty=true]{4.78(22)}{\kilo\pascal} & \num[separate-uncertainty=true]{3.1(1)} \\
		Protein gels & \SI[separate-uncertainty=true]{2.42(19)}{\kilo\pascal} & \num[separate-uncertainty=true]{2.7(1)}\\
		\hline
	\end{tabular}
	\label{Tab:PowerLaw}
\end{table}

Alternatively, if any emulsion gel is seen as a protein-filled droplet gel matrix, then the reinforcing role of the proteins can be expressed by \[\frac{G'^{~exp}_{mixture}}{G'^{~model}_{droplet}(\phi_{droplet})}\] where the storage modulus of the matrix $G'^{~model}_{droplet}$ is also calculated using the characterisation of pure droplet gels as a function of the volume fraction.

\begin{figure}[htbp]
	\begin{center}
		\includegraphics[width=0.95\textwidth]{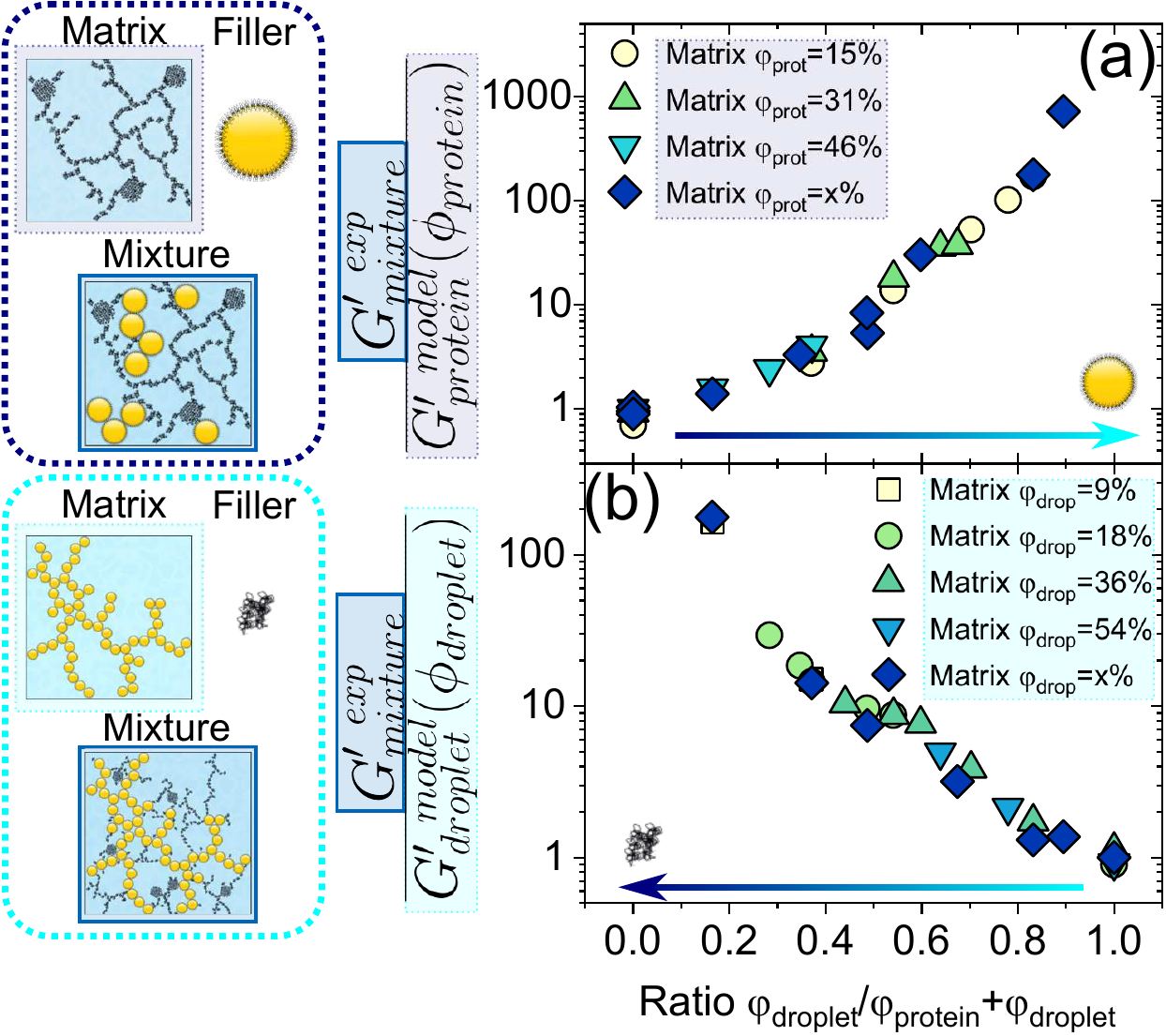}
	\end{center}
	\caption{Reinforcement of a protein gel upon addition of droplets  $G'^{~exp}_{mixture}/ G'^{~model}_{protein}(\phi_{protein})$ (top, from left to right), and of a droplet gel upon addition of proteins $G'^{~exp}_{mixture}/G'^{~model}_{droplet}(\phi_{droplet})$ (bottom, from right to left) as a function of the relative amount of droplet added $\phi_{eff, drop}/(\phi_{eff, prot}+\phi_{eff, drop})$. The miscellaneous volume fractions of matrices are indicated by x\%, and the values can be found in Table S1. The two graphs represent the same samples of gels of mixtures, as shown in Figure~\ref{Fig:GelMixtModuli}, but differ by the arbitrary role of the components: the proteins form the matrix in the top graph while they are the fillers in the bottom graph, and vice-versa for the droplets, as depicted in the cartoon.
	}
	\label{Fig:GelMixtReinforc}
\end{figure}

The two scenarios, droplet-filled protein gels and protein-filled droplet gels, are used for the analysis of the gels of mixtures presented in Figure~\ref{Fig:GelMixtModuli}, and the reinforcement in both cases is shown in Figure~\ref{Fig:GelMixtReinforc}. The reinforcement of the gel is represented as a function of the proportion of droplets in the mixture, rather than the volume fraction of droplets, in order to facilitate comparisons of gels at different concentrations

As can be seen, there is a collapse of the reinforcing effects for matrices of different volume fraction to a single master curve in both cases.  For the two scenarios, the elastic modulus is doubled when the amount of filler added is  $\SI{25}{\percent}$ of the matrix volume fraction (\textit{i.e.} $\phi_{filler}/(\phi_{filler}+\phi_{matrix})=\num{0.2}$), and grows ten-fold when the amount of filler is equal to the volume fraction of the matrix (\textit{i.e.} $\phi_{filler}/(\phi_{filler}+\phi_{matrix})=\num{0.5}$). The increase in storage modulus as a function of the relative amount of fillers is thus independent of the density of the matrix.

This invariability is probably related to the fractal nature of the colloidal gels studied here. Indeed, for all the gels of mixtures, the matrix, whether protein gel or droplet gel, is a fractal gel with a heterogeneous structure, as illustrated in the first part of this study \cite{roullet:2020a}. For the fillers to significantly reinforce this structure, they must contribute to the network as much as the particles forming the matrix gel and their amount has thus to be  calculated relative to the matrix density rather than in absolute terms, in which case the master curve does not appear.

Furthermore, either scenario of matrix/filler pairs gives a similar result, which seems to indicate that protein-coated droplets and un-adsorbed proteins have a symmetric contribution to the viscoelasticity of the gels of their mixtures. The ability of the two components to form a gel of their own may be the source of behaviour. Hence, the established approach of emulsion gels as droplet-filled protein gels does not reflect the complex structure of these systems when the droplets are small enough. Instead of matrix and fillers, it may thus be more appropriate to consider emulsion gels as fractal composite networks made of both proteins and droplets.

\subsubsection{Intermediate behaviour of the composite networks: influence of the composition}

In this second approach to the viscoelastic behaviour of emulsion gels, they are envisioned as composite colloidal gels of total volume fraction \\ $\phi_{eff,prot}+\phi_{eff,drop}$ and for which the composition indicates how similar they are to pure gels of droplets and of proteins. The focus is thus moved from the reinforcement of a matrix with the addition of another colloidal species, to the comparison of the composite networks with pure gels at the same total volume fraction.

The storage modulus $G'^{~exp}_{mixture}$ of the gels formed by the mixtures can be compared to the weighted mean of the storage moduli of the gels formed by their pure components. This can be achieved using the power law dependence on the volume fraction identified in the first part of this study for pure gels \cite{roullet:2020a}. A theoretical storage modulus $G'^{~model}_{mixture}$ for the emulsion gels can thus be expressed by a linear rule of mixture:
\begin{eqnarray}
G'^{~model}_{mixture} &=& \frac{\phi_{droplet}}{\phi_{protein}+\phi_{droplet}}\times G'^{~model}_{droplet}(\phi_{protein}+\phi_{droplet}) \nonumber \\ & &  + \left(1-\frac{\phi_{droplet}}{\phi_{protein}+\phi_{droplet}}\right)\times G'^{~model}_{protein}(\phi_{protein}+\phi_{droplet})
\label{Eq:StorModMixtModel}
\end{eqnarray} 
Where $G'^{~model}_{protein}$ and $G'^{~model}_{droplet}$ designate the modulus of a hypothetical protein gel (resp. droplet gel), containing the same total volume fraction $\phi_{protein}+\phi_{droplet}$ as the mixture, and calculated using Equation~\ref{Eq:PureGels} and the values presented in Table~\ref{Tab:PowerLaw}.

\begin{figure}[htbp]
	\begin{center}
		\includegraphics[width=0.8\textwidth]{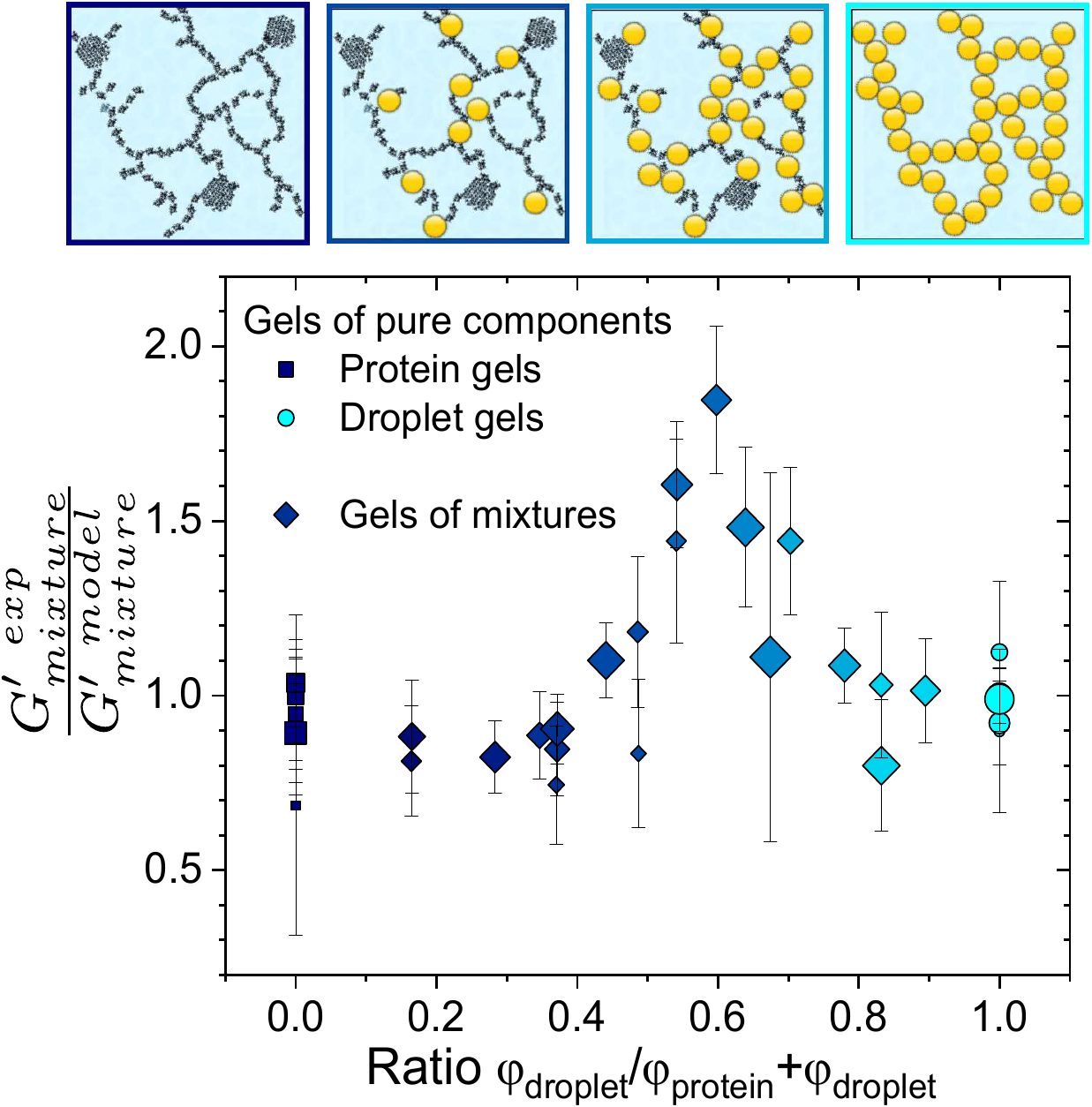}
	\end{center}
	\caption{Ratio between the experimental storage modulus $G'^{~exp}_{mixture}$ and the theoretical storage modulus $G'^{~model}_{mixture}$, defined in Equation~\ref{Eq:StorModMixtModel} as the weighted mean of the pure gels moduli, as a function of the relative amount of droplets $\phi_{eff, drop}/(\phi_{eff, prot}+\phi_{eff, drop})$ illustrated in the cartoon.  The size of the data points indicates the total volume fraction $\phi_{eff, drop}+\phi_{eff, prot}$. This graph represents the same gel samples than shown in Figure~\ref{Fig:GelMixtModuli}. The error bars arise from error propagation upon calculation of the theoretical storage modulus, and take into account the errors of the models for each pure component.
	}
	\label{Fig:DistToProtMixt}
\end{figure}

The ratio between experimental and theoretical storage moduli\\ $G'^{~exp}_{mixture}/G'^{~model}_{mixture}$ is shown in Figure~\ref{Fig:DistToProtMixt} as a function of the composition, described by the ratio $\phi_{eff, drop}/(\phi_{eff, prot}+\phi_{eff, drop})$. Schematically, this figure can be interpreted as the change in gel strength in a network of fixed volume fraction when its composition goes from a pure protein gel to a pure droplet gel.

As can be seen, Equation~\ref{Eq:StorModMixtModel} provides a good approximation of the storage modulus of emulsion gels over a large part of the composition range, as the ratio $G'^{~exp}_{mixture}/G'^{~model}_{mixture}\approx 1$. A noticeable increase of this ratio is observed for the mixtures with $\SI{50}{\percent}<\phi_{eff, drop}/(\phi_{eff, prot}+\phi_{eff, drop})<\SI{70}{\percent}$, for which the experimental storage modulus is moderately higher than the weighted mean of the pure gels. Apart from this minor deviation, the storage moduli of gels made of droplets and proteins follow the rule of mixture that is generally used for composite materials \cite{ashby:2011-11}, where the fraction of each component is given by the proportion of the total volume fraction of the system.

In addition, this dependence of the mixtures storage moduli as a function of the composition does not apparently depend on the total volume fraction, indicated by the size of the data points in Figure~\ref{Fig:DistToProtMixt}. Such a result seems to imply that the two compositional parameters of the framework introduced earlier can be decoupled, and their contribution to the properties of the mixtures can be analysed separately.

Finally, a consequence of this decoupling and of the characterisation of the influence of the ratio of components is the ability to estimate the storage modulus of an emulsion gel of known composition. Indeed, a rough estimation of its strength can first be calculated from the power law identified for a protein gel in the first part of this study, using its total volume fraction $\phi_{eff, drop}+\phi_{eff, prot}$. It is then corrected for the composition of the mixture by using the trend for the normalised storage moduli displayed in Figure~\ref{Fig:DistToProtMixt}.

\subsection{Frequency dependence of emulsion gels}

\begin{figure}[htbp]
	\begin{center}
		\includegraphics[width=0.55\textwidth]{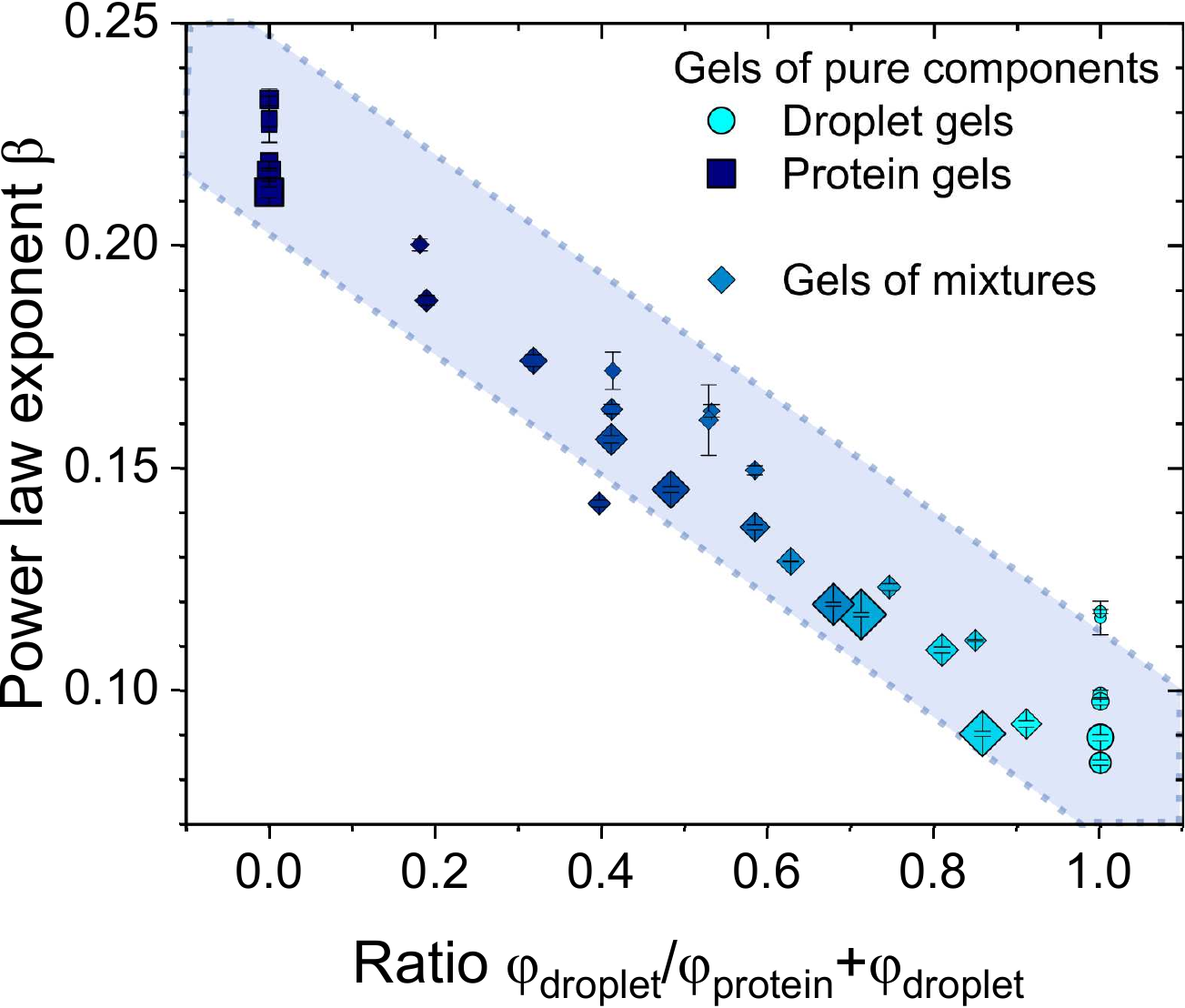}
	\end{center}
	\caption{Comparison of frequency dependence for gels of mixtures (diamonds, colour-coded), of protein stabilised droplets (circles, in cyan) and of protein (squares, in dark blue): power-law exponent $\beta$ as a function of the ratio $\phi_{eff,drop}/(\phi_{eff, prot}+\phi_{eff, drop})$. The size of the data points indicates the total volume fraction $\phi_{eff, drop}+\phi_{eff, prot}$. The shaded area is a guide for the eye.
	}
	\label{Fig:FreqExpMixt}
\end{figure}

Similarly to the pure protein and droplet gels presented in the first part of this study, the frequency dependence of emulsion gels was measured after gelation, and is represented in Figure S3 of the supplementary material \cite{roullet:2020a}.
The dependence of the storage modulus of emulsion gels on frequency can be modelled by a power law, as was done for pure gels: \[G'=G'_{0,\omega}\times \omega^{\beta} \] Where the exponent $\beta$ describes the dynamic behaviour of the networks. Here $\beta$ is estimated for each emulsion gel and presented as a function of the composition in Figure~\ref{Fig:FreqExpMixt}.

This representation of the frequency dependence of the network as a function of the ratio $\phi_{eff,drop}/(\phi_{eff, prot}+\phi_{eff, drop})$ demonstrates that there is a continuous transition between that of droplet gels, at the lower end of the horizontal axis, and of protein gels, at the upper end of the horizontal axis. Indeed, the frequency dependence of mixtures presents some variations with the total volume fraction, represented by the size of the data points, but varies overall between $\beta_{droplet}\approx\num{0.1}$ and  $\beta_{protein}\approx\num{0.2}$ as the proportion of protein increases in the mixture. This is in good correspondence with previous studies in which a decrease in frequency dependence was observed upon addition of casein-coated droplets in a casein gel \cite{vanvliet:1988}. 

Therefore, it seems that the difference in dynamics between droplets and proteins is reflected linearly in the mixtures as a function of their composition. 
This result reinforces the hypothesis that emulsion gels are composite networks that are best described as intermediate between protein gels and droplet gels. 

\section{Conclusion}

The choice of the parameters used for the description of protein-stabilised emulsion gels with sub-micron droplets is the first step in giving shape to a new framework for these systems. Here, based on qualitative arguments about the structure of colloidal gels, it is suggested that this category of soft solids can be viewed as composite networks made of droplets and protein assemblies. The composition of these systems was thus defined by their total volume fraction $\phi_{eff, prot}+\phi_{eff, drop}$ and composition ratio $\phi_{eff,drop}/(\phi_{eff, prot}+\phi_{eff, drop})$. These parameters were calculated by using a previous study on the viscosity of pure suspensions of proteins and of droplets \cite{roullet:2019a}. This two dimensional composition range of emulsion gels was explored in this study.

The analysis of the rheological properties of emulsion gels in this framework confirmed the relevance of this choice. Indeed, it was found that the storage modulus is mostly determined by the total volume fraction of the emulsion gel $\phi_{eff, prot}+\phi_{eff, drop}$. In addition, when the strength of the emulsion gels is scaled in order to account for the variations in volume fraction, it varies continuously between the behaviour of pure protein gels and pure droplet gels following a simple rule of mixture. Similarly, the frequency dependence varies continuously  between the behaviour of protein gels and droplet gels, linearly with the composition ratio $\phi_{eff,drop}/(\phi_{eff, prot}+\phi_{eff, drop})$. Notably, the decoupling of total volume fraction and relative composition for the rheological properties justifies \textit{a posteriori} the choice of parameters.

In addition, the viscoelasticity of the emulsion gels presented here was also analysed using the classical approach of droplet-filled matrix \cite{dickinson:2012,farjami:2019}. It was shown that the total volume fraction is more important than the absolute amount of fillers, as the reinforcing effect of fillers collapsed onto a mastercurve when scaled by the density of the matrix. This finding shines a new light on previous studies of the rheology of attractive droplet-filled emulsion gels \cite{vanvliet:1988,chen:1999b,dickinson:1999,koenig:2002,sala:2007,gravelle:2015}. In addition, the symmetric role of the components may reinforce the idea of composite networks, where the stress-bearing strands are formed by the proteins and protein-stabilised droplets alike. The classical approach for these systems thus yields results that seem to reinforce the image of protein-stabilised emulsion gels as intermediate colloidal gels.


The implications of these results are multiple. A first obvious application is the formulation of dairy products with a more refined control of their rheological properties, as the present study offers a more precise characterisation of the contributions of un-adsorbed proteins and of sub-micron droplets. This falls within the emerging framework of dairy products, like milk and cheese, envisioned as soft colloidal systems \cite{roullet:2019a,gillies:2019}.

More generally, the description of emulsion gels as intermediate colloidal gels could offer a model for the formulation of emulsion gels of fine-tuned rheology. Indeed, the study of emulsion gels of any composition could be performed in two steps. First, pure gels of the two components are characterised over a wide range of volume fraction, in what could be described as a calibration step. Then, using the quantification of the variation in the intermediate zone between the two limit systems, the properties of any gel of mixture can be calculated using their composition. Such an analytical approach to formulation would present the advantage of identifying a small range of possible composition to reach the required mechanical properties, rather than using a more common ``trial and error'' process.

Finally, in a broader picture, mixture systems are not commonly studied in academic research, despite being ubiquitous in industrial products. Here a simple framework for thinking about emulsion gels is suggested. In this, they are first deconstructed into their components, protein-stabilised droplets and un-adsorbed proteins, and then compared to these primary systems. This approach may be valid for a larger range of ternary mixtures in which two components play a similar role in building up the viscoelasticity, while the solvent plays none. Further investigations are needed to identify other systems that can be modelled as composite networks.

\section{Acknowledgements}
This project forms part of the Marie Curie European Training Network COLLDENSE that has received funding from the European Union’s Horizon 2020 research and innovation programme Marie Sk\l{}odowska-Curie Actions under the grant agreement No. 642774

\appendix





\bibliographystyle{model1-num-names}
\bibliography{References}







\end{document}